# New phenomena beyond Spontaneous Symmetry Breaking

**Y. Contoyiannis[1], P. Papadopoulos[1], L. Matiadou[1],**
**S.G. Stavrinides[2], M. Hanias[2], S.M. Potirakis[1]**
[1]*Electrical and Electronic Engineering Department, University of West Attica, Athens, Greece*
[2]*Physics Department, International Hellenic University, Kavala, Greece*

***Abstract:*** It is known that thermal systems of finite size that are subject to second-order phase transitions and until the Spontaneous Symmetry Breaking (SSB) is completed, the fluctuations of the order parameter obey in the dynamics of critical intermittency [6]. Beyond the SSB, critical intermittency doesn't hold; consequently, it is not expected that the distribution of the waiting times in the order parameter timeseries would hold any power law.
However, we reveal for the first time that right after the SSB power laws still exist within a small zone of temperatures. These power laws emerge due to another form of intermittency that determines the dynamics of the order parameter fluctuations in the beginning of a tri-critical crossover, without this crossover ever being completed in a first-order phase transition.
In the work presented hereby, we present and explain this change of the dynamics of the order parameter fluctuations, as the temperature drops under the temperature of SSB. Finally, it is mentioned that such a phenomenon has been already observed in pre-seismic processes.

## 1. Introduction

It has been found that in any finite thermal physical system exhibiting a second-order phase transition according to the $\phi^4$ theory [1-3], a temperature zone emerges, below the critical temperature, which could be considered as pseudocritical in finite size systems [4-7]. An example of such a system is the classic paradigm of the 3D-Ising spin system.
Inside this zone, the second-order phase transition critical point remains, even though the mathematical symmetry in the Landau free energy is broken. This region between the critical (in fact pseudocritical) temperature $T_c$ and the temperature $T_{SSB}$ ($T_{SSB}$<$T_c$), where the critical point appears for the last time, is called the hysteresis zone and its properties have been studied in detail in works of ours [4,6].
The purpose of the present work is to investigate what happens in a region very close to the $T_{SSB}$, just below the SSB temperature. In specific, we investigated whether the scaling laws were still valid according to the dynamics of critical intermittency or if these dynamics of order parameter fluctuations had been changed. The relevant results prove the emergence of new dynamics in the zone just below the SSB. Finally, an interpretation explaining this alteration in the dynamics is provided.

## 2. The 3D-Ising model beyond the SSB

Introducing the essential notation and theoretical backgraound, in the case of a *Z(N)* spin system, spin variables are defined as: $s(a_i) = e^{i2\pi a_i/N}$ (lattice vertices $i = 1 \dots i_{max}$) with $a_i = 0,1,2,3 \dots N-1$. The well-known Ising models correspond to the case of *N*=2. On the other hand, Metropolis is an effective algorithm producing configurations and in this algorithm the configurations at constant temperatures are selected with Boltzmann statistical weights $e^{-\beta H}$, where *H* stands for the Hamiltonian of the spin system; than the nearest neighbors' interactions can be written as:

$$\mathrm{H} = -\sum_{<i,j>} \mathrm{J_{ij} s_i s_j}. \qquad (1)$$

As known [1,3], according to this model a second-order phase transition takes place, when the temperature drops below a critical value. Thus, for a $32^3$ lattice in three dimensions (3D-Ising model), the critical (or pseudocritical for finite size lattices) temperature has been found to be $T_c = 4.515$ ($J_{ij}$ =1) [6]. Considering that a sweep of the whole lattice represents the algorithmic time unit and that the possible spin values are ±1, then the numerically calculated mean magnetization $M$ timeseries, which forms the order parameter, demonstrates a trajectory of fluctuations of this order parameter.

According to a previous work of ours [6], it has been that in the case of a lattice $32^3$ the SSB temperature has been calculated to be $T_{SSB}$=4.45 ($J_{ij}$ =1). In Figure 1 the distribution of the magnetization at SSB temperatures for 300.000 configurations is presented. It is apparent that the separation of the two lobes of the distribution has been completed.

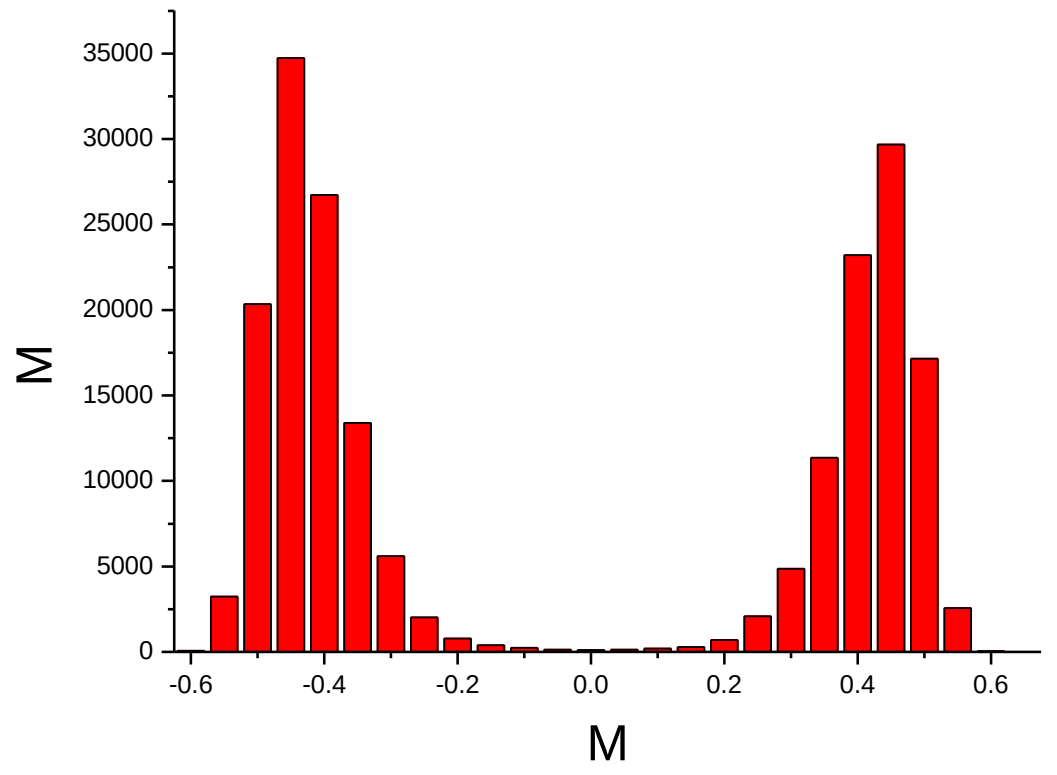

***Figure 1.*** *The distribution of mean magnetization in the* $T_{SSB}$*=4.45 (The separation of the two lobes is completed).*

In Figure 2 the magnetization distributions right after the SSB (with a second decimal approximation) at T=4.44, as well as at a lower temperature T= 4.2, are presented. As one may see, for temperatures right after the SSB, as in Figure 2(a), the distribution is not symmetric but shows a shift of the maximum towards higher values and an extension of the tail, towards smaller values. But as we get further away from SSB, as in Figure 2(b),the distribution tends to become symmetrical.

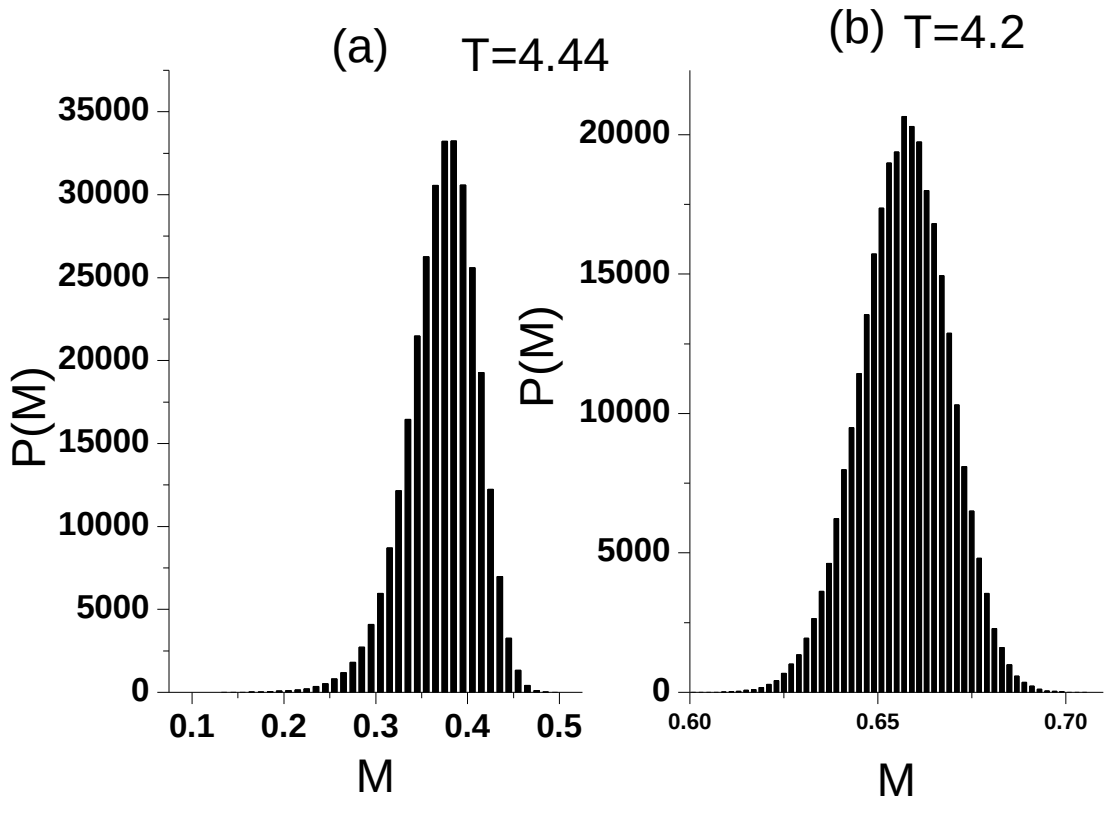

***Figure 2.*** *(a) The distribution of the mean magnetization right after the SSB (second decimal approximation) at T=4.44. (b) The distribution of the mean magnetization at T=4.2.*

The dynamics of the critical state in equilibrium conditions are the dynamics of intermittency. Mathematically, these are expressed through the Type I intermittent map (see equation 4 below) [8,9]. The phenomenon of intermittency consists of alternating regions of small fluctuations, known as laminar regions, interrupted by chaotic bursts, and the waiting times within the laminar regions are called laminar lengths. In the case of critical intermittency, the laminar length ($L$) distributions obey to a power law of the form [9,10]:

$$P(L) \sim L^{-p} \tag{2}$$

with the condition that [9]:

$$p = 1 + \frac{1}{\delta} \tag{3}$$

where $\delta$ is one of the six critical exponents [1]; for thermal systems it is called the isothermal critical exponent. Given that $\delta$>1, it is expected that the critical exponent would within $p \in$ (1,2) [1]. Thus, finding power laws for the distribution of laminar lengths with exponent $p$ in the above interval, indicate the existence of criticality.

## 3. Data Analysis

In [11] the novel Method of Critical Fluctuations (MCF) has been introduced to reveal the existence of critical dynamics in a time series we have introduced [11] the method of critical fluctuations (MCF). This method is applied in the case of the time series of the mean magnetization at temperature *T*=4.44, which corresponds to the distribution appearing in Figure 2(a). Utilizing this distribution, one may define the as the laminar region the one that is lies within the edge corresponding to the peak $M_0$ (fixed point) and the edge passing from a point $M_l$ at the less steep region of the distribution. We consider the position of $M_l$ as a free parameter, corresponding to that value for which a distribution gets closer to the power law. Then, the laminar lengths are calculated as the waiting times of the values of the mean magnetization in the interval $[M_0, M_l]$. In the case of the distribution in Figure 2 (a), such an interval is [0.45, 0.32] for which the following distribution of laminar lengths is obtained (Figure 3).

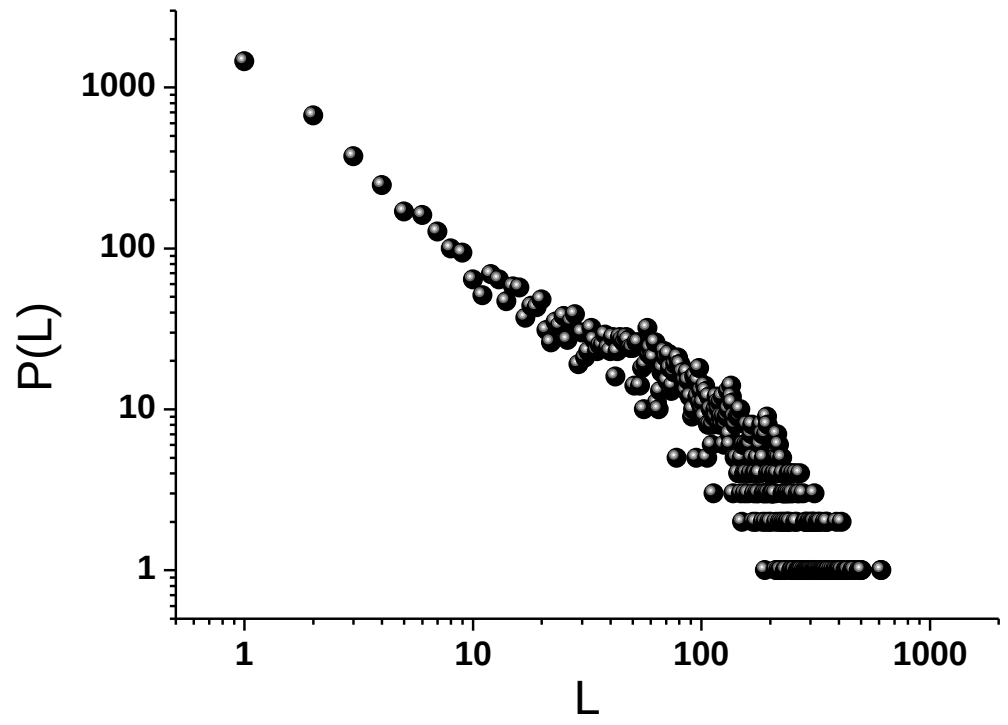

***Figure 3.*** *The laminar length distribution for the mean magnetization at temperature T=4.44.*

The fitting function for calculating the critical exponents $p1$ and $p1$ is usually of the form $\mathrm{f(x) = p_1 x^{-p_2} e^{-p_3 x}}$. However, in this form case and due to the change of curvature from a point and on, the previous relation is not a suitable function for fitting the experimental points, unless we consider only the small scales and disregard the larger ones. Moreover, it is known that any power-law having the form of eq. (2) is valid for *L*>>1 [10-16]. Thus, we need a method capable of revealing the power laws for larger scales, where the classical least squares methods do not provide reliable results due to the small statistics of the points in the tails of the distributions.

In [17] such a method has been introduced. According to this, all points on all scales are considered, especially those in great scales, ignoring the noise that usually appears in the tails of such distributions. This method results from the development on a Haar Wavelet basis and reveals the power law, if it exists, no matter how strong the noise. The wavelet base is a linear base suitable for phenomena that exhibit self-similar properties, such as critical phenomena. For the sake of clarity, this method and its algorithm are presented in the appendix of this paper, while in the main text we provide directly the results. We strongly advise the reader to read the Appendix before proceeding to the rest of the main text

In Figure 4, we show the last ten values for parameter λ, according to wavelet analysis, which show the distance from the unit. The calculation of the λ-values is performed according to equation (A2). Following the second step of the algorithm quantity $D_\lambda$ is calculated (eq. 4A).

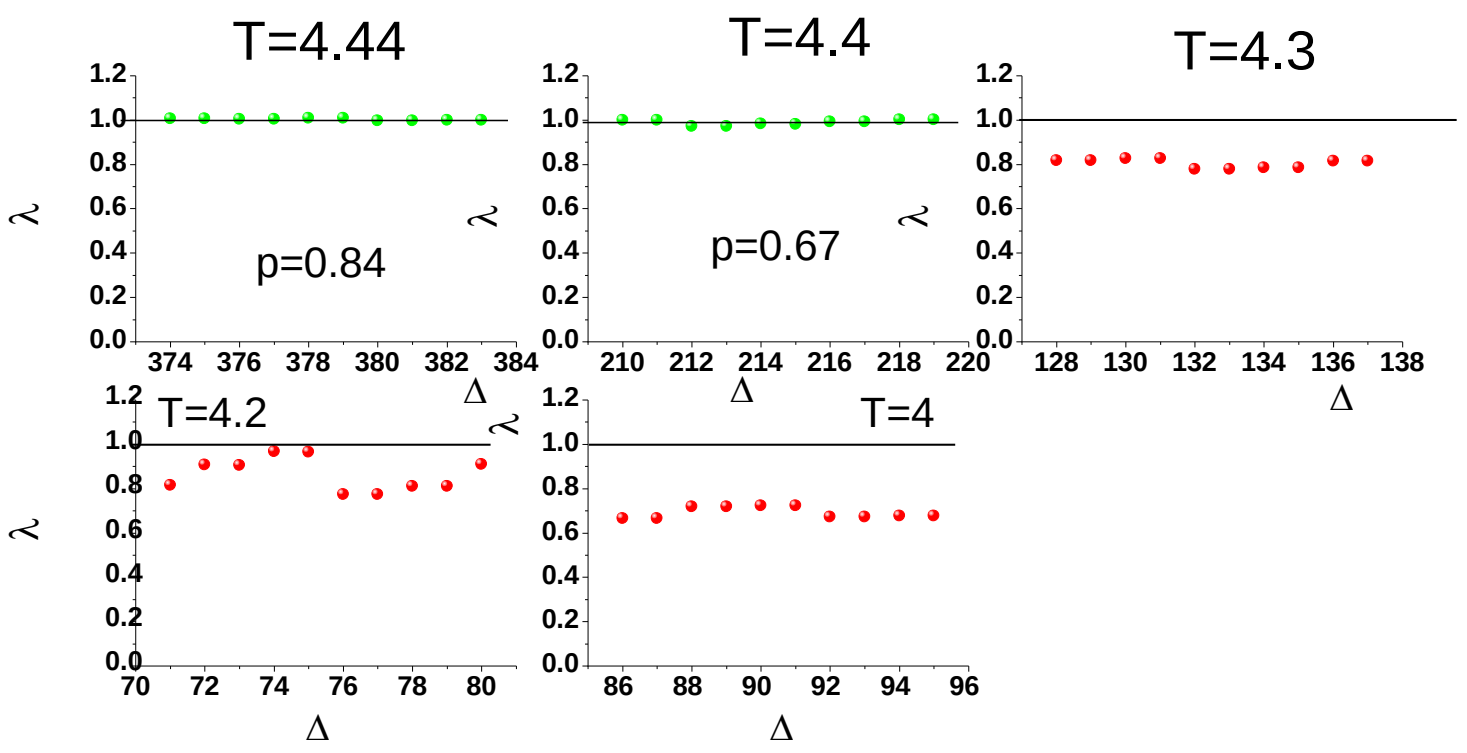

***Figure 4.*** *For temperatures T=4.44 and T=4.4 the calculated exponents are 0.84 and 0.67 respectively. For those below T=4.4, no exponent can be calculated since we have moved away from the power laws.*

This way a quantitative measure, estimating the proximity to the power law, is provided. The closer the quantity $D_\lambda$ gets to zero, the closer we are to the ideal power law. We mention the rule of thumb that holds: the ideal power law, i.e., when $\lambda$ converges to 1, occurs when $D_\lambda < 4\cdot10^{-4}$. In TABLE 1 we present the results of wavelet-based analysis for the temperatures appearing in Figure 4.

**TABLE 1**

| Temperature T | Dλ | $\Delta_{max}$ | Power-Law | Exponent p |
|---|---|---|---|---|
| 4.44 | $4.3 \cdot 10^{-5}$ | 334 | YES | 0.84 |
| 4.4 | $2.23 \cdot 10^{-4}$ | 219 | YES | 0.67 |
| 4.3 | $3.8 \cdot 10^{-2}$ | 137 | NO | - |
| 4.2 | $8.5 \cdot 10^{-2}$ | 80 | NO | - |
| 4 | $9.56 \cdot 10^{-2}$ | 95 | NO | - |

## 4. The zone beyond SSB

As already mentioned, critical dynamics are mathematically expressed by the Type I intermittency map, which bears the following form [9]:

$$M_{n+1} = M_n + uM_n{}^z + \varepsilon_{n,} \qquad (4)$$

with $\varepsilon_n$ representing the added noise necessary for the ergodicity of the system [18]. The exponent *z* of the nonlinear term is related to the isothermal critical exponent $\delta$ through the relation $z=\delta+1$ [9]. If the dynamics described in eq. (4) are applied on the system when this is within the hysteresis zone, i.e. $T_{SSB}< T <T_{cr}$ [4,6], then the exponents *p* should have values in the interval [1,2). Looking at Figure 4, it is apparent that the exponent values are bounded within the interval [0.66, 1), leading to the reasonable quest for the reason of the existence of these values?

In [19], the existence of intermittent dynamics obeying to a different map, compared to the one of eq. (4) was presented and the relevant dynamics studied. This map had the following form:

$$M_{n+1} = M_n - vM_n{}^{-z} + \varepsilon_n, \tag{5}$$

where the generated distribution of laminar lengths demonstrates an exponent provided by the following relation:

$$p = \frac{z}{z+1} = \frac{\delta+1}{\delta+2}. \tag{6}$$

Provided that the isothermal critical exponent gets values in the interval (1,∞) it follows that p∈[0.66,1). Within these dynamics the order parameter at the beginning of a process of metastable states, known as tricritical crossover, ends in the first-order phase transition, studied by $\phi^6$ Landau theory [19]. One characteristic feature of the specific dynamics is that the fixed point of the map shifts from zero (eq. 4) to infinity (eq. 5).

In realistic systems of finite size, the fixed point gets finite values. In Figure 5, we present the distribution of *M,* when the intermittent map of eq. (5) is applied, for setting the values: $z = 4,\ v = 0.2, \varepsilon_n \in [-0.01,0.01], fixed\ point = 3, number\ iterations = 100.000$ .

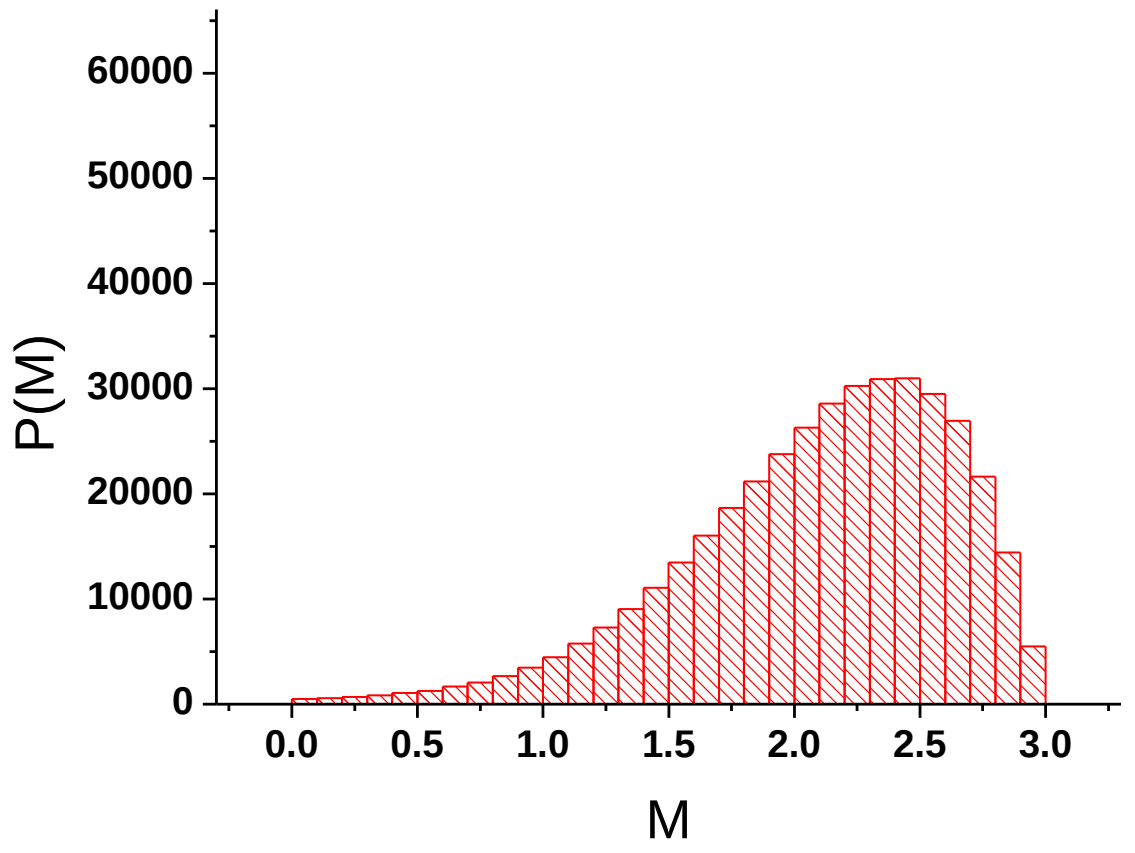

***Figure 5.*** *The distribution of M values produced by the intermittent map in eq. (5).*

Comparing Figures 5 and 2(a), one realizes that they share common characteristic features and in particular an asymmetry with respect to the peak, as it has shifted to higher values, while an extension of the tail at small values to almost zero is present.

As a consequence of the above, one may claim that right after the SSB and while the temperature is decreased, the dynamics of the order parameter fluctuations are no longer determined by the critical intermittency of eq. (1); on the contrary, now the order parameter fluctuations are determined by the intermittency dynamics of eq. (5), which are the dynamics of the beginning of the tricritical crossover.

To quantitatively determine the extent of this temperature zone, we produce the $D_\lambda$ vs T diagram, which appears in Figure 6. As mentioned above, the quantity $D_\lambda$, is a quantitative criterion expressing the proximity of the relevant dynamics to the power law that emerged according to the analysis with the wavelet method.

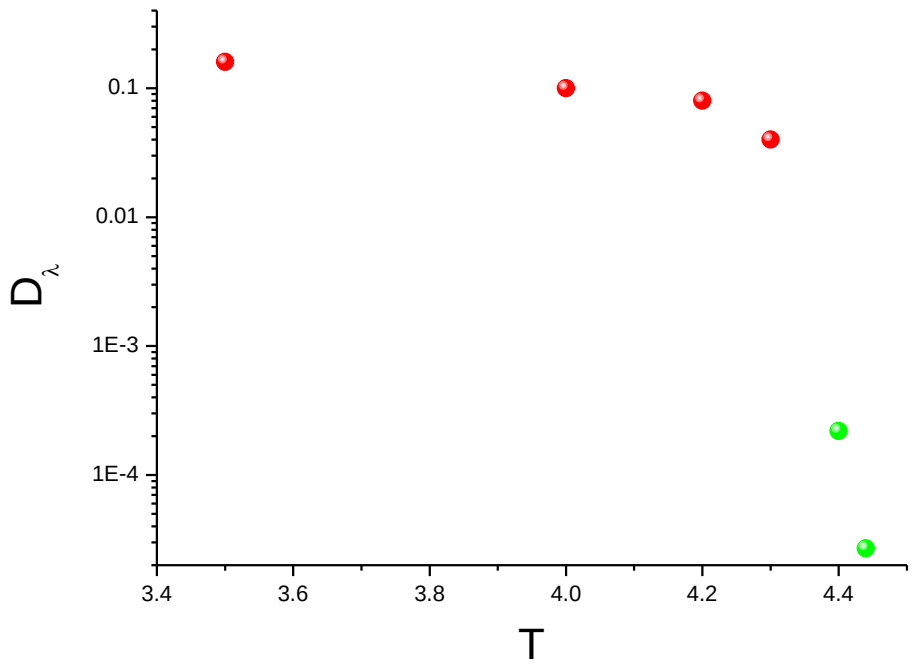

***Figure 6.*** *The $D_\lambda$ vs T diagram. The green points belong to the zone of eq. (5) dynamics, and the red ones are beyond this zone at lower temperatures. We see a sharp transition from the zone (green) where we find power laws for the distribution of laminar lengths, to the interval (1,0.66] at temperatures T<4.4 where the distribution of laminar lengths deviates from the power-law.*

Thus, according to our study, right after the SSB, a very narrow zone of temperatures exists, where the power laws are maintained but the exponents lie within the interval [0.66, 1) i.e. below unity. This zone ranges from $T$=4.44 to $T$=4.4.

The above could be confirmed by setting $\delta$=4.8, which is the value of the isotherm critical exponent of the universality class of the 3D-Ising model [20], and calculating (eq. 6) the theoretical value of exponent $p$, which becomes $p$=0.85. This value is very close to the one we calculated numerically in this work for the 3D-Ising model at temperature T=4.44, which is $p$=0.84 according to the Wavelet Method (see table 1).

## 5. Discussion

It is known that the Z(2) spin theory is a $\phi^4$ theory and it does not provide the existence of a tricritical point. Considering the results presented in the lines above a very serious challenge emerges and this is the interpretation of the existence of this very narrow temperature zone right after the SSB.

Towards an interpretation, we remind that right after the SSB (Figure 2a) the unstable critical point, where the mean magnetization is zero, ceases to exist as a fixed point and is shifted from zero value to finite values. Since the second-order phase transition of the $\phi^4$ theory is a continuous phase transition and not an abrupt one, the dynamics of intermittency that existed until the SSB, continue to exist for continuity reasons. The difference is that the the fixed point will no longer have a zero value; instead, the new fixed points are now placed at higher values. As we have thoroughly discussed in [19], the region where the intermittent dynamics develop could be determined in a Landau free-energy diagram; in this case they are determined by surfaces that present a plateau. Looking for such regions in the Landau free-energy diagram we present their form throughout the evolution of the studied phenomenon in Figure 7.

In Figure 7 (a), at the beginning of the tricritical crossover to the first-order phase transition according to the theory$\phi^6$, two small plateaus appear, and the corresponding fixed points are not placed at zero value, but at the edges of these regions (the walls of the potential). This is because, after the break in symmetry and due to the continuity of the phase transition, the system needs to maintain the intermittent behavior through the new phase, as well. This justifies the existence of the tri-critical dynamics right after the SSB, and consequently the existence of power laws in the range of values [0.66,1). After a very narrow temperature zone these plateaus vanish.

However, the system does not evolve according to the $\phi^6$ theory, as Figures 7(b) and 7 (c) show. This is because the $\phi^6$ theory predicts the existence of 3 fixed points, with the one in the middle being the initial fixed point at zero (Figure 7b and 7c). Since in $\phi^4$ theory, this fixed point ceases from being a fixed point after the SSB, this theoretical framework becomes limited only to its beginning. We could say that the $\varphi^6$ theory " has lent" the tricritical point for a very small temperature width, in the $\phi^4$ theory for reasons of continuity of the dynamics of intermittency in a new form.

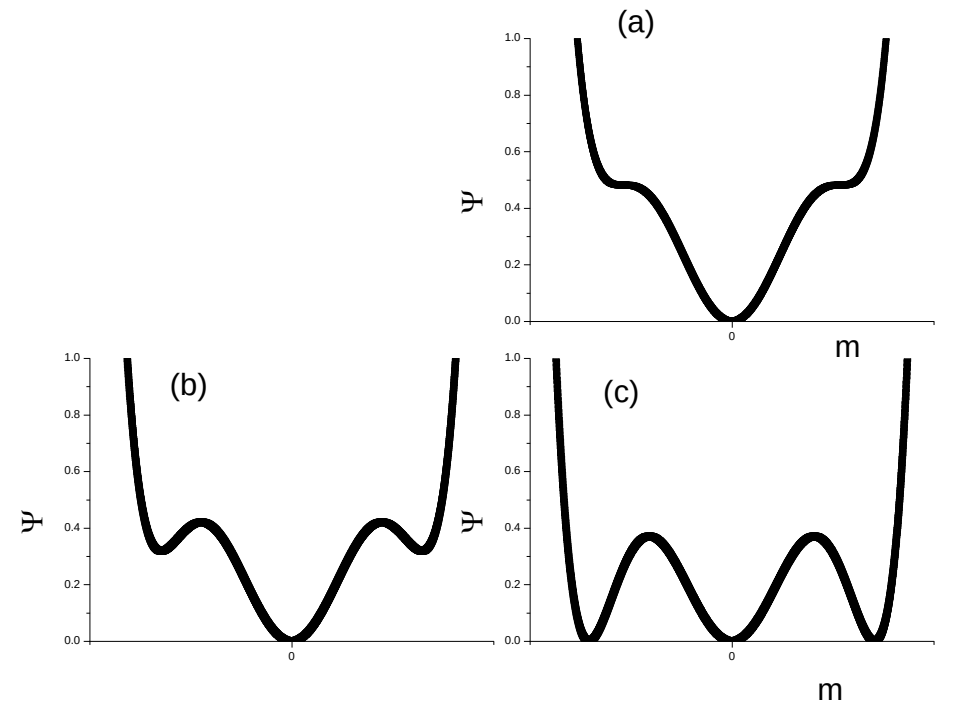

***Figure 7.*** *The Landau free energy Ψ vs the order parameter m. (a) The beginning of the tricritical crossover. (b) An intermediate metastable state, where the stable fixed points of the $\phi^6$ theory appear. (c) The degeneracy of the three fixed points. It is the point of the first order phase transition.*

All that we have presented in this section are absolutely compatible as we will say in the following, with the theory of the tricritical point as it is presented in references [1,21]. According to the theory, in the parametric diagram of the two coefficients in the first terms $\varphi^2$ and $\varphi^4$ of the Landau free energy there is the point (0,0) where the two parameters are becoming zero and thus the lines of the second and first order phase transitions meet. This point is the tricritical (according to Griffiths [1]) point.

What we reveal in present work is that there is really the temperature of the SSB where the two dynamics meet, i.e. the critical intermittency (eq. 4) and the tri-critical intermittency (eq. 5). Thus, the SSB the dynamics of the second-order phase transition ends and the first-order phase transition begins. Of course, as we mentioned, due to the fact that the symmetries of $\varphi^4$, $\varphi^6$ theories, i.e. Z(2) and Z(3) respectively, are different, only the beginning of the tri-critical crossover appears. So we come to the conclusion that in the meeting of the lines of the phase transitions as the theory provide in the parametric diagram corresponds to the "coupling" in SSB between the critical intermittency and the tri-critical intermittency for the fluctuations of the order parameter.

## 6. Conclusions

In thermal systems of finite size that undergo a second-order phase transition according to the $\varphi^4$ theory and after the SSB is completed, a narrow temperature zone appears, with the particular characteristic that the scaling laws for the distribution of waiting times are maintained, with exponents being in the range $p \in [0.66,1)$. The dynamics of the order parameter fluctuations within this zone don't follow the Type I critical intermittency dynamics, which characterize the second-order phase transition; instead, another form of intermittency determining the dynamics of the order parameter fluctuations at the beginning of a tricritical crossover, appears. It is interesting to investigate how strong this narrow-band phenomenon can become, so that it could be detectable in real systems, as well as the consequences caused. Hinting applications of these unusual dynamics have appeared in pre-seismic processes. Such applications will be the subject of future work.

## Appendix

Recently we have introduced a new method of analysis [17], when we will have to take into account large scales as well. This method results from the development on a wavelet basis and reveals the power law, if it exists, no matter how strong the noise. The wavelet base is a linear base suitable for phenomena that exhibit self-similar properties such as critical phenomena. The wavelet undergo two transformations, the change of scale j and their displacement k. Thus, the coefficients of the analysis are $d_{j,k}$. When j = k = 0 we have the coarse graining description of the analysis.In the framework of this description, the coefficients of the analysis can and do ignore the noise of the signal to be analyzed [17]. We use this behavior to develop an algorithm that applies to each distinct numerical or real signal f (i), i = 1,…$\Delta_{max}$,with the $\Delta_{max}$ the maximum length of the signal. In practice, $\Delta_{max}$ will be found in the tail of the distribution as the point for which we will find the most ideal power law (step 2 of the algorithm that follows).This algorithm is able to answer the question of whether a signal is a power law, how close or far it is from the power law, and it can also calculates the corresponding exponent p. In other words, it plays the role ofa fitting function without carrying the pathogenicity of the fitting function due to noise, especially at the high values of the laminar lengths. The new method uses all scales. The base we use to develop the algorithm is the Haar wavelet base. The Haar base has, as mother function, the following function defined by the Theta-functions for space [0, Δ]:

$$\psi_H = \Theta\left(\frac{\Delta}{2} - x\right)\Theta(x - 0) - \Theta\left(x - \frac{\Delta}{2}\right)\Theta(\Delta - x). \tag{1A}$$

We define the quantities [17]:

$$\lambda = \frac{\frac{d_{00}}{d_{10}}}{\frac{d_{10}}{d_{20}}} = \frac{d_{00}d_{20}}{d_{10}^2} = \frac{\left(\sum_{i=1}^{\frac{\Delta}{2}} f(i) - \sum_{\frac{\Delta}{2}}^{\Delta} f(i)\right)\left(\sum_{i=1}^{\frac{\Delta}{8}} f(i) - \sum_{\frac{\Delta}{8}}^{\frac{\Delta}{4}} f(i)\right)}{\left(\sum_{i=1}^{\frac{\Delta}{4}} f(i) - \sum_{\frac{\Delta}{4}}^{\frac{\Delta}{2}} f(i)\right)^2} \tag{2A}$$

and

$$R = \frac{d_{00}}{d_{10}} = \frac{1}{\sqrt{2}}\left(\sum_{i=1}^{\frac{\Delta}{2}} f(i) - \sum_{\frac{\Delta}{2}}^{\Delta} f(i)\right) / \left(\sum_{i=1}^{\frac{\Delta}{4}} f(i) - \sum_{\frac{\Delta}{4}}^{\frac{\Delta}{2}} f(i)\right). \tag{3A}$$

The proposed method for revealing the criticality and finding the exponent of the power law of distribution of laminar lengths has the following steps:

1. We apply the equation (2A) to calculate λ as a function of Δ up to Δmax. As we can see from (2A) the minimum Δ that can give information is Δ = 8. We make the plot λ vs Δ and because we are interested in the convergence of λ [17] the last 10 (10>8) points are enough to deduce conclusion.
2. We quantify the previous step by calculating the distance of λ from thevalue λ = 1 which is the perfect power law [17], by calculating the quantity

$$D_\lambda = \frac{1}{10}\sum_{i=\Delta_{max}-9}^{\Delta_{max}} (1 - \lambda_i)^2 \tag{4A}$$

Obviously as the $D_\lambda$ is closer to zero the distribution of laminar lengths will be closer to a power-law. This is the criterion that determines $\Delta_{max}$.

3. From equation (3A) we produce the plot R vs Δ. From the convergence regionofthe diagram ($\leq 10$)an average value for the quantity R is obtained.
4. We consider $f(i) = ci^{-p}$ , $i = 1,2,3 \ldots \Delta_{max}$ as a test function in (3A) (c does not matter because in (3A) it is vanished) and by solving numerically the equation (3A) we calculate the exponent p for R closer to the average value which we found in step 3.

To create the programs for calculating the above quantities, we use the discrete form of $d_{j,k}$ according to the relation: [17]

$$d_{j,k} = \sqrt{\frac{2^j}{\Delta}}\left(\sum_{i=\max\left(\left[\frac{k\Delta}{2^j}\right],1\right)}^{\left[k\frac{\Delta}{2^j}+\frac{\Delta}{2^{j+1}}\right]} f(i) - \sum_{i=\left[k\frac{\Delta}{2^j}+\frac{\Delta}{2^{j+1}}\right]+1}^{\left[(k+1)\frac{\Delta}{2^j}\right]} f(i),\right. \tag{A5}$$

where [z] means the integer part of a variable z.